\documentclass[aps,prd,twocolumn,amsmath,superscriptaddress,floatfix,nofootinbib,preprintnumbers]{revtex4-2}
\usepackage{amssymb}\usepackage{txfonts}\usepackage{epsfig}\usepackage{bm}
\usepackage{color}\usepackage{graphicx,graphics}\usepackage{multirow}\usepackage{float}\usepackage{ulem}
\usepackage{setspace}\usepackage{slashed}\usepackage{booktabs}\usepackage{hyperref}
\hypersetup{colorlinks=true, citecolor=blue, linkcolor=red, filecolor=black,urlcolor=blue}
\allowdisplaybreaks[4]

\begin{document}

\title{Hadron production in the charged current semi-inclusive deeply inelastic scattering of $N=Z$ nuclei}

\author{Wenyan Yu}
\author{Weihua Yang}

\affiliation{Department of Nuclear Physics, Yantai University, Yantai, Shandong 264005, China}


\author{Xing-hua Yang}
\affiliation{School of Physics and Optoelectronic Engineering, Shandong University of Technology, Zibo, Shandong 255000, China}

\begin{abstract}
 The charged current weak interaction can distinguish quark flavors, it provides a valid method to determine (transverse momentum dependent) parton distribution functions in high energy reactions by utilizing tagged hadrons. In this paper, we calculate the charged current semi-inclusive deeply inelastic neutrino and anti-neutrino scattering of $N=Z$ nuclei. Semi-inclusive means that a spin-1 hadron is also measured in addition to the scattered charged lepton. The target nucleus has the same number of neutrons and protons and is assumed as unpolarized.  According to calculations, we find that only chiral-even terms survive and chiral-odd terms vanish in the differential cross section for this charged current deeply inelastic (anti-)neutrino nucleus scattering process. Furthermore, we introduce a measurable quantity, the yield asymmetry of the produced hadron $A^h$, to determine the nuclear parton distribution functions. Numerical estimates show that the yield asymmetry is independent of the type of target nucleus if it has the same number of neutrons and protons. 
 Numerical estimates also show that  sea quark distribution functions and disfavored fragmentation functions have significant influence on measurable quantities.

\end{abstract}

\maketitle

\section{Introduction}\label{sec:introduction}

Parton distribution functions (PDFs) and fragmentation functions (FFs) are important quantities in describing high energy reactions. The former gives information of the nucleon or nucleus structures while the latter gives information about the hadronization process. The one-dimensional PDFs and FFs are usually measured in the inclusive deeply inelastic scattering (DIS) process and the inclusive electron positron annihilation process, respectively. Nevertheless, the three-dimensional or transverse momentum dependent (TMD) PDFs and FFs can be determined simultaneously in the hadron production semi-inclusive DIS (SIDIS) process. In addition to the charged lepton DIS of nucleon, the deeply inelastic neutrino nucleus scattering has attracted renewed attention in studying the nucleon and nucleus structures \cite{Feng:2017uoz,FASER:2018bac,Anchordoqui:2021ghd,Feng:2022inv}. In the past three decades, various experiments have confirmed the fact that PDFs measured in free nucleons and nuclei in which nucleons are bound are significantly different. This implies that nuclei are not simple accumulations of protons and neutrons and they would have non-nucleon degrees of freedom \cite{Frankfurt:1988nt}. Fortunately, (anti-)neutrino nucleus scattering not only provides information on the flavor separation which cannot be realized in the charged lepton DIS experiments alone but also can be used to probe the nuclear modification effects and to determine nuclear parton distribution functions (nPDFs) with different nuclear targets. For example, the neutrino nucleus scattering can be used to study the EMC effect at the valence quark kinematic region \cite{EuropeanMuon:1983wih}. 
Beyond that, it can also provide information of the electro-weak interaction by calculating the neutral current and charged current DIS processes. For example, they  can provides precision measurements of the weak mixing angle \cite{Boer:2011fh} by measuring the Paschos-Wolfenstein ratio \cite{Paschos:1972kj}. 


Inclusive DIS can only access one-dimensional nucleon structures or information about one-dimensional PDFs, as mentioned before. Inclusive means that only the scattered lepton is detected in the final state. Semi-inclusive DIS can access three-dimensional nucleon structures or TMD PDFs. Semi-inclusive implies that a jet or a hadron in the current fragmentation region is also detected in addition to the scattered lepton in the final state. For the jet production SIDIS, jet is a direct probe of analyzing properties of the quark transverse momentum. In the familiar $\gamma^* N$ collinear frame, for example, the transverse momentum of the virtual photon ($\vec q_\perp$) is zero, and therefore the transverse momentum of the jet ($\vec k'_\perp$) is equal to that of the incident quark ($\vec k_\perp$) if the higher order gluon radiations are neglected \cite{Yang:2022xwy,Yang:2023vyv,Yang:2023zod,Wu:2023omf}. More information about jet production SIDIS for PDFs studies can be found in Refs. \cite{Song:2010pf,Song:2013sja,Wei:2016far,Gutierrez-Reyes:2018qez,Gutierrez-Reyes:2019vbx,Liu:2018trl,
Liu:2020dct,Kang:2020fka,Arratia:2020ssx,Chen:2020ugq,Yang:2020qsk}. Nevertheless, jet production SIDIS process can not cover the low energy kinematic regions and access the chiral-odd PDFs due to the conservation of the helicity \cite{Yang:2022sbz}. Worse yet, jet production SIDIS has difficulties to identify light flavor PDFs  since no hadron is measured or tagged. Under this condition, one has to consider the hadron production SIDIS process \cite{Mulders:1995dh,Bacchetta:2006tn} where PDFs and FFs are measured simultaneously with flavor identifications. 

To meet the needs of future high-energy neutrino experiments (e.g., FASER \cite{Feng:2017uoz}),  in this paper, we calculate the charged current semi-inclusive deeply inelastic neutrino and anti-neutrino nucleus scattering of $N=Z$ nuclei. The target nucleus has the same number of neutrons and protons and is assumed as unpolarized. In the inclusive process, the formalism of the neutrino nucleus scattering is the same to that of the neutrino nucleon scattering \cite{SajjadAthar:2022pjt}. We assume that this conclusion still holds in the semi-inclusive process. As a result, the differential cross section of the semi-inclusive process would be given in terms of convolutions of nuclear PDFs and FFs. Hadronization is also influenced by nuclear medium, but we do not consider the nuclear modified FFs here. Since charged current weak interaction can only couple to the left-handed fermions or helicities do not flip, contributions to differential cross sections from chiral-even terms survive and that from chiral-odd terms vanish. In the following section, we will show that only $f_1$ contributes actually.  
In order to determine $f_1$ in the neutrino and anti-neutrino nucleus scattering, we introduce a measurable quantity,  the yield asymmetry of the produced hadrons $A^h$. Here $h$ can be pion, kaon and other hadrons. Numerical estimates show that the yield asymmetry is independent of the type of target nucleus provided it has the same number of neutrons and protons. In other words, yield asymmetry is not sensitive to the type of  $N=Z$ nuclei. It therefore serve as a good candidate to determine the nuclear distribution functions. Furthermore, we notice that sea quark distribution functions and disfavored FFs have significant influence on measurable quantities.  We also provide a set of the azimuthal asymmetries in this paper.

To be explicit, we organize this paper as follows. In Sec. \ref{sec:sidis}, we present the formalism of the charged current semi-inclusive deeply inelastic neutrino and anti-neutrino scattering.  In order to calculate the cross section, we choose the $VA$ collinear frame where $V$ denotes $W$ boson and the target particle travels in the positive $z$ direction. Conventions used in this paper are also given. In Sec. \ref{sec:hadronic}, we calculate the hadronic tensor and the differential cross section in the parton model.  Discussions of  measurable quantities and numerical estimates are presented in Sec. \ref{sec:yielda}. Finally, we present a brief summary in Sec. \ref{sec:summary}.

\section{The Formalism}\label{sec:sidis}

\begin{figure}
  \centering
  \includegraphics[width=0.5\linewidth]{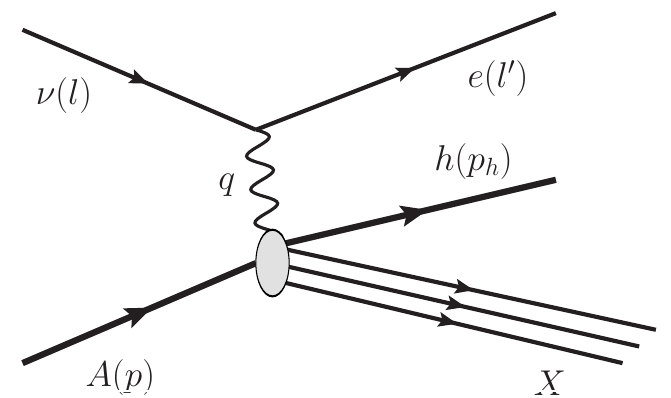}
  \caption{The Feynman diagram of the charged current (anti-)neutrino nucleus SIDIS process. Momenta of the incident particles are given in parentheses.}\label{fig:nuFeyn}
\end{figure}

We consider the charged current (anti-)neutrino nucleus SIDIS process where a spin-1 hadron rather than a jet is measured in addition to the scattered charged lepton in the final state, see Fig. \ref{fig:nuFeyn}. To be explicit, we label this SIDIS process in the following form,
\begin{align}
  \nu(l) + A(p) \rightarrow e(l^\prime) + h(p_h, S) + X,
\end{align}
where $\nu$ and $e$ denote the incoming (anti-)neutrino and the outgoing charged lepton with momenta $l$ and $l'$, respectively. $A$ with momentum $p$ denotes the unpolarized target nucleus with the same number of neutrons and protons. $h$ is the produced hadron with momenta $p_h$ and spin $S$. We assume that the formalism of the SIDIS of nucleus is the same to that of nucleon and the standard variables are defined in the same way, 
\begin{align}
  & s=(p+l)^2, \quad \quad  x=\frac{Q^2}{2 p\cdot q},  \quad \quad   y=\frac{p\cdot q}{p \cdot l},\nonumber\\
  & z=\frac{p\cdot p_h}{p\cdot q}=-\frac{2p_h\cdot q}{Q^2},  \label{f:sidisvar}
\end{align}
where $Q^2=-q^2=-(l-l')^2$. 

The differential cross sections of the neutrino nucleus scattering and the anti-neutrino nucleus scattering can be written as a unified form, 
\begin{align}
  d\sigma = \frac{\alpha_{\rm em}^2}{2sQ^4}A_W L_{\mu\nu}(l, l^\prime)W^{\mu\nu}(p,p_h)\frac{d^3 \vec l^\prime }{E^\prime} \frac{d^3\vec p_h}{E_h}, \label{f:crosssec}
\end{align}
where  $\alpha_{\rm em}$ is the fine structure constant,
\begin{align}
  A_W=\frac{Q^4}{\left[(Q^2+M_W^2)^2+\Gamma_W^2M_W^2\right]16\sin^4\theta_W}.
\end{align}
Here $M_W, \Gamma_W$ are the mass and width of $W$ boson, $\theta_W$ is the weak mixing angle. The kinematic factor can be written in terms of Fermi constant $G_F$. In this case, Eq. (\ref{f:crosssec}) is written as 
\begin{align}
  d\sigma = \frac{G_F^2}{16s\pi^2} \frac{M_W^4}{(Q^2+M_W^2)^2} L_{\mu\nu}(l, l^\prime)W^{\mu\nu}(p,p_h)\frac{d^3 \vec l^\prime }{E^\prime} \frac{d^3\vec p_h}{E_h}. \label{f:crosssecgf}
\end{align}
The leptonic tensor is given by
\begin{align}
  L_{\mu\nu}(l,l')=2\left(l_\mu l'_\nu+l_\nu l'_\mu -g_{\mu\nu} l\cdot l'\right)+2i\lambda_\nu \varepsilon_{\mu\nu l l'}, \label{f:leptonicten}
\end{align}
where $\lambda_\nu$ is introduced for convenience. For neutrino,  $\lambda_\nu$ is $-1$ and for anti-neutrino $\lambda_\nu$ is $1$. The hadronic tensor is given by
\begin{align}
  W^{\mu\nu}=&\int\frac{d^4k }{(2\pi)^4}\frac{d^4k'}{(2\pi)^4}\delta^4(k'-k-q){\rm{Tr}}\left[\Phi(k)\Gamma^\mu\Xi(k')\Gamma^\nu\right], \label{f:hadronictensor}
\end{align}
where distribution correlator $\Phi(k)$ and fragmentation correlator $\Xi(k')$ are respectively defined as
\begin{align}
  \Phi(k) &=\int d^4\xi e^{ik\xi}\langle A |\bar{\psi}(0)\psi(\xi)|A\rangle, \label{f:Phi} \\
  \Xi(k') &=\sum_{X}\int d^4\zeta e^{ik'\zeta}\langle 0 |\psi(0)|h,X\rangle\langle h, X|\bar{\psi}(\zeta)|0\rangle. \label{f:Xi}
\end{align}
Here $k$ and $k'$ are momenta of the incident quark in the nucleus and the scattered quark fragmenting into hadrons. 
We have neglected gauge links for short notations. $\Gamma^\mu=\gamma^\mu(c_V^q-c_A^q\gamma^5)$ and $c_V^q, c_A^q$ are weak couplings. For the charged current weak interaction considered in this paper, $c_V^q=c_A^q=1$.

\begin{figure}
  \centering
  \includegraphics[width=0.65\linewidth]{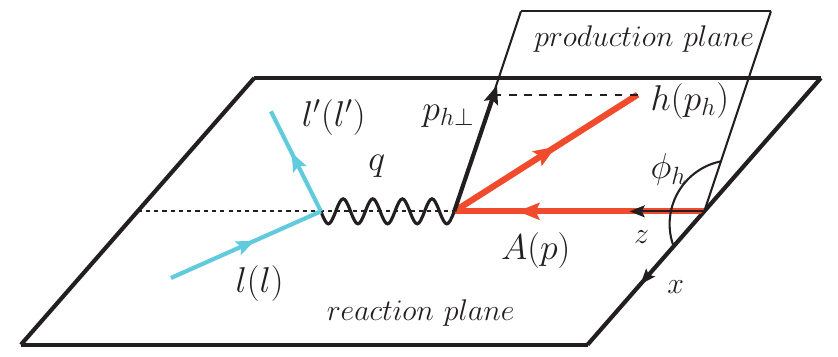}
  \caption{Illustration of the SIDIS process in the $VN$ collinear frame. The target nucleus travels in the positive $z$ direction. $\phi_h$ is the angle between hadron production plane and the reaction plane. The former is determined by momenta of the produced hadron and the nucleus while the latter is determined by momenta of the neutrino and the nucleus.
  }\label{fig:nusidis}
\end{figure}

In order to calculate the hadronic tensor and the cross section, we choose the $VA$ collinear frame, see Fig. \ref{fig:nusidis}, the target nucleus travels in the positive $z$ direction. The x-o-z plane is just the reaction plane which is determined by the momenta of the neutrino and the nucleus. Here $V$ denotes the gauge boson, in this paper it is $W$ boson. 
We introduce two unit vectors  $\bar n^\mu=\frac{1}{\sqrt{2}}(1, 0, 0, 1)$, $n^\mu=\frac{1}{\sqrt{2}}(1, 0, 0, -1)$.
They satisfy $\bar n^2=n^2=0$ and $\bar n\cdot n=1$. In light-cone coordinates, $\bar n^\mu=(1, 0, \vec 0_\perp),  n^\mu=(0,  1, \vec 0_\perp)$. Therefore the momenta of the target nucleus and the gauge boson can be parameterized as 
\begin{align}
  & p^\mu= p^+ \bar n^\mu + p^- n^\mu, \label{f:ptildemudef}\\
  & q^\mu =q^+ \bar n^\mu + q^- n^\mu. \label{f:qmudef}
\end{align}
Nucleus usually has a large mass. However, in this paper we do not consider the mass effect. Under this condition, only the large plus component of $p^\mu$ survives up to $\mathcal{O}(1/Q^2)$ level.  Light-cone vectors can be defined as $\bar n^\mu=p^\mu/p^+, n^\mu=(q^\mu+xp^\mu)/q^-$ and momenta related to this hadron production SIDIS process take the following forms,
\begin{align}
  p^\mu &=\left(p^+, 0, \vec{0}_\perp \right), \label{f:pplus}\\
  q^\mu &=\left(-xp^+, \frac{Q^2}{2xp^+},  \vec{0}_\perp \right), \\
  l^\mu &=\left(\frac{1-y}{y}xp^+, \frac{Q^2}{2xyp^+}, \frac{Q\sqrt{1-y}}{y}, 0 \right), \\
  l^{\prime \mu}&=\left(\frac{1}{y}xp^+, \frac{(1-y)Q^2}{2xyp^+}, \frac{Q\sqrt{1-y}}{y}, 0 \right), \\
  p_h^\mu &=\left(\frac{\vec{p}^2_{h\perp}}{zQ^2}xp^+, \frac{zQ^2}{2xp^+}, \vec{p}_{h\perp} \right), \\
  p^\mu_{h\perp} &=|\vec{p}_{h\perp}|\left(0, 0, \cos\phi_h, \sin\phi_h \right). \label{f:momenta}
\end{align}
From these definitions, the four-dimensional delta function in the hadronic tensor can be decomposed as 
\begin{align}
  \delta^4(k'-k-q)&=\delta^+(k^{\prime +}-k^+-q^+)\delta^-(k^{\prime -}-k^- -q^-)\nonumber\\
  &\times \delta^2(\vec{k}_\perp^\prime - \vec{k}_\perp-\vec{q}_\perp).
\end{align}
The hadronic tensor therefore is reduced as
\begin{align}
  \hat W^{\mu\nu}&=\int d^2 \vec k_\perp d^2 \vec k_\perp^{\prime}\delta^2(\vec{k}^\prime_\perp- \vec k_\perp-\vec q_\perp) \nonumber \\
   &\times {\rm{Tr}}\left[\hat \Phi(x,k_\perp)\Gamma^\mu\hat \Xi(z,k'_\perp)\Gamma^\nu\right], \label{f:hadronicint}
\end{align}
where the three-dimensional distribution correlator and fragmentation correlator are given by
\begin{align}
  \hat\Phi(x,k_\perp) &=\int \frac{d\xi^-d^2\vec\xi_\perp}{(2\pi)^3} e^{ixp^+\xi^- -i\vec{k}_\perp\vec{\xi}_\perp }\nonumber\\
  &\times \langle A |\bar{\psi}(0)\psi(\xi^-,\vec{\xi}_\perp)|A\rangle, \label{f:Phiint} \\
  \hat \Xi(z,k'_\perp) &=\sum_{X_h}\int \frac{d\zeta^+d^2\vec\zeta_\perp}{(2\pi)^3} e^{ip_h^-\zeta^+/z -i\vec{k^\prime}_\perp\vec{\zeta}_\perp } \nonumber\\
  &\times\langle 0 |\psi(0)|h,X\rangle\langle h, X|\bar{\psi}(\zeta^+,\vec{\zeta}_\perp)|0\rangle. \label{f:Xiint}
\end{align}

By using the following relationships $d^3\vec{p}_h/E_h=dzd^2\vec p_{h_\perp}/z$ and $d^3\vec l/E'=sydxdyd\psi/2$, $\psi$ is the azimuthal angle of $\vec l^\prime$ around $\vec l$, we can rewrite the differential cross section in the following form,
\begin{align}
  \frac{d\sigma}{dxdyd\psi dz d^2\vec{p}_{h_\perp}}=\frac{y\alpha^2_{\rm em}}{4zQ^4}A_WL_{\mu\nu}\hat W^{\mu\nu}. \label{f:crosssection}
\end{align}

One can calculate the hadronic tensor and the cross section in terms of structure functions, see Ref. \cite{Bacchetta:2006tn}. However, we here only calculate them in the parton model. Our calculations are limited at the leading-twist level.

\section{Calculations in the parton model}\label{sec:hadronic}

\subsection{Hadronic tensor in the parton model}

Both the distribution correlator and the fragmentation correlator shown in Eqs. (\ref{f:Phiint}) and (\ref{f:Xiint}) are $4\times 4$ matrices in Dirac space and can be decomposed in terms of the Dirac Gamma-matrices and coefficient functions, 
\begin{align}
  &\hat{\Phi}=\frac{1}{2}\left(\Phi I +\tilde{\Phi}\gamma^5+\Phi_\alpha \gamma^\alpha +\tilde{\Phi}_\alpha\gamma^\alpha\gamma^5 +\Phi_{\alpha\beta} i\sigma^{\alpha \beta}\gamma^5\right),  \label{f:hatphi}\\
  &\hat{\Xi}=\frac{1}{2}\left(\Xi I +\tilde{\Xi}\gamma^5+\Xi_\alpha \gamma^\alpha +\tilde{\Xi}_\alpha\gamma^\alpha\gamma^5 +\Xi_{\alpha\beta} i\sigma^{\alpha \beta}\gamma^5\right). \label{f:hatxi}
\end{align}
The coefficient functions can be further decomposed in terms of nuclear PDFs and FFs. For the coefficient functions of the distribution correlator, only two terms survive at leading twist after the decomposition, they are, 
\begin{align}
  &\Phi^\alpha=\bar{n}^\alpha f_1, \label{f:pdff1}\\
  &\Phi^{\alpha_\beta}=\bar{n}^\alpha \frac{\varepsilon_\perp^{k \beta}}{M_N} h^\perp_1. 
\end{align}
For the coefficient functions of the fragmentation correlator, we have
\begin{align}
   z\Xi^\alpha =& n^\alpha \Bigg(D_1+\frac{\varepsilon_\perp^{k^\prime S}}{M_h} D^\perp_{1T}+ S_{LL}D_{1LL} \nonumber\\
  &+\frac{k_\perp^\prime \cdot S_{LT}}{M_h}D^\perp_{1LT} +\frac{S^{k'k'}_{TT}}{M_h^2}D^\perp_{1TT}\Bigg), \\ 
   z\tilde{\Xi}^\alpha =& n^\alpha \Bigg(\lambda_h G_{1L}+\frac{k_\perp^\prime \cdot S_{T}}{M_h} G^\perp_{1T}\nonumber\\
  &+\frac{\varepsilon_\perp^{k^\prime S_{LT}}}{M_h}G^\perp_{1LT} +\frac{\tilde{S}^{k'k'}_{TT}}{M_h^2}G^\perp_{1TT}\Bigg), \\ 
  z\Xi_{\alpha\beta}=& n^\alpha \Bigg[\frac{\varepsilon_\perp^{k^\prime \beta}}{M_h} \left(H^\perp_1 +S_{1LL}H^\perp_{1LL}\right)+ S_T^\beta H_{1T}\nonumber\\
  &+ \frac{k_\perp^{\prime \beta}}{M_h} \left(\lambda_h H^\perp_{1L} + \frac{k_\perp^\prime \cdot S_{T}}{M_h}H^\perp_{1T}\right) \nonumber \\
  &+\varepsilon_\perp^{S_{LT}\beta} H_{1LT} + \frac{\tilde{S}^{k'k'}_{TT}}{M_h^2}H^{\prime\perp}_{1TT}\nonumber\\
  &+ \frac{\varepsilon_\perp^{k^\prime \beta}}{M_h} \left(\frac{k_\perp^\prime \cdot S_{LT}}{M_h}H^\perp_{1LT}+\frac{S^{k'k'}_{TT}}{M_h^2}H^{\perp}_{1TT}\right) \Bigg]. \label{f:zxialphabeta}
\end{align}
Lorentz-scalar quantities $M_N$ and  $M_h$ are introduced via dimensional analysis.  $\varepsilon_\perp^{\rho\mu}k_{\perp\rho}=\varepsilon^{\rho\mu\alpha\beta}\bar{n}_\alpha n_\beta k_{\perp\rho}$ and $\tilde{S}^{k'k'}_{TT}=\varepsilon_{\perp \alpha}^{k'}S^{\alpha k'}_{TT}$. $\lambda_h$ is the helicity of the produced hadron, $S_T$ is the transverse polarization vector and can be parametrized as
\begin{align}
& S_T^\mu = |S_T| \left( 0,0, \cos\phi_S, \sin\phi_S \right). \label{f:Smu}
\end{align}
For the tensor polarization dependent parameters, we define and parameterize them as in Ref.~\cite{Bacchetta:2000jk},
\begin{align}
& S_{LT}^x = |S_{LT}| \cos\phi_{LT}, \\
& S_{LT}^y = |S_{LT}| \sin\phi_{LT}, \\
& |S_{LT}| = \sqrt{(S_{LT}^x)^2 + (S_{LT}^y)^2}, \\
& S_{TT}^{xx} = -S_{TT}^{yy} = |S_{TT}| \cos2\phi_{TT}, \\
& S_{TT}^{xy} = S_{TT}^{yx} = |S_{TT}| \sin2\phi_{TT}, \\
& |S_{TT}| = \sqrt{(S_{TT}^{xx})^2 + (S_{TT}^{xy})^2}.
\end{align}

In order to calculate  the hadronic tensor, we would use the following traces,
\begin{align}
 & {\rm{Tr}}\left[\slashed{\bar{n}}\Gamma^\mu\slashed{n}\Gamma^\nu\right] ={\rm{Tr}}\left[\slashed{\bar{n}}\gamma^5\Gamma^\mu\slashed{n}\gamma^5\Gamma^\nu\right]=-4c_1^qg_\perp^{\mu\nu}-4ic_3^q\varepsilon_\perp^{\mu\nu}, \\
 & {\rm{Tr}}\left[\slashed{\bar{n}}\gamma^5\Gamma^\mu\slashed{n}\Gamma^\nu\right] ={\rm{Tr}}\left[\slashed{\bar{n}}\Gamma^\mu\slashed{n}\gamma^5\Gamma^\nu\right]
  =-4c_3^qg_\perp^{\mu\nu}-4ic_1^q\varepsilon_\perp^{\mu\nu}, \\
 & {\rm{Tr}}\left[i\sigma^{-\alpha}\gamma^5\Gamma^\mu i\sigma^{+\beta}\Gamma^\nu\right]=4c_2^q\left(g_\perp^{\mu\nu}g_\perp^{\alpha\beta} -g_\perp^{\mu\alpha}g_\perp^{\nu\beta}-g_\perp^{\mu\beta}g_\perp^{\nu\alpha}\right). \label{f:traceh}
\end{align}
Here $g_\perp^{\mu\nu}=g^{\mu\nu}-\bar n^\mu n^\nu -\bar{n}^\nu n^\mu$, $\varepsilon_\perp^{\mu\nu}=\varepsilon^{\mu\nu\alpha\beta}\bar{n}_\alpha n_\beta $. These weak interaction factors are defined as 
\begin{align}
  & c_1^q=(c_V^q)^2+(c_A^q)^2,\\
  & c_2^q=(c_V^q)^2-(c_A^q)^2, \label{f:c2q}\\
  & c_3^q=2c_V^q c_A^q.
\end{align}
Finally, we obtain the hadronic tensor which is given in terms of $f_1$ and corresponding  FFs, 
\begin{align}
 \hat W^{\mu\nu}&=-\frac{1}{z}\int d^2 \vec k_\perp d^2 \vec k_\perp^{\prime}\delta^2(\vec{q}_\perp +\vec{k}_{\perp} -\vec{k}^{\prime}_{\perp})\nonumber\\
 &\times \left[\left(c_1^q g_\perp^{\mu\nu}+i c_3^q \varepsilon_\perp^{\mu\nu}\right) f_1\mathcal{D}
+\left(c_3^q g_\perp^{\mu\nu}+i c_1^q \varepsilon_\perp^{\mu\nu}\right)f_1\mathcal{G}\right]. \label{f:hadronicTen}
\end{align}
We have use the short-handed notations, 
\begin{align}
  \mathcal{D}&=\mathcal{D}_1+\frac{\varepsilon_\perp^{k^\prime S}}{M_h} D^\perp_{1T} +\frac{k_\perp^\prime \cdot S_{LT}}{M_h}D^\perp_{1LT} +\frac{S^{k'k'}_{TT}}{M_h^2}D^\perp_{1TT}, \\
  \mathcal{G}&=\lambda_h G_{1L}+\frac{k_\perp^\prime \cdot S_{T}}{M_h} G^\perp_{1T}+\frac{\varepsilon_\perp^{k^\prime S_{LT}}}{M_h}G^\perp_{1LT} +\frac{\tilde{S}^{k'k'}_{TT}}{M_h^2}G^\perp_{1TT},
\end{align}
where $\mathcal{D}_1=D_1+ S_{LL}D_{1LL}$. 
From Eq. (\ref{f:hadronicTen}) we see that chiral-odd terms vanish and only chiral-even terms survive. It is easily to understand this conclusion. In calculating the chiral-odd terms we have used the trace in Eq. (\ref{f:traceh}), in which a factor $c_2^q$ is obtained. In the charged current weak interaction considered in this paper, the weak interaction couplings $c_V^q=c_A^q=1$. As a result $c_2^q=0$ and the coupled chiral-odd terms vanish naturally. From a physical perspective, $W$ boson only couple to left-handed fermions.  Helicities of these quarks do not flip and chiral symmetries are conserved. Under this constraint, chiral-odd terms must vanish.

From Eq. (\ref{f:hadronicTen}), we also notice that the hadronic tensor depends on transverse momenta of the incident quark and scattered quark, $k_\perp$ and $k_\perp^\prime$. They can not be measured in the experiments. In order to calculate the cross section, it is best expressed in terms of measurable quantities defined via the transverse momentum of the produced hadron rather than that of relevant quarks. To this end, we introduce a unit vector defined via the transverse momentum of the produced hadron, $h^\mu=p_{h\perp}^\mu/|\vec{p}_{h\perp}|$. The transverse momentum of the quark ($k_\perp^\mu$) is therefore replaced by $-(h\cdot k_\perp)h^\mu$ and the cross section is expressed in terms of measurable quantities of the produced hadron. One can find further introductions and applications to $h^\mu$ in Ref. \cite{Tangerman:1994eh}. Under this framework, the hadronic tensor can be rewritten as 
\begin{align}
 \hat W^{\mu\nu}&=-\frac{1}{z}\int d^2 \vec k_\perp d^2 \vec k_\perp^{\prime}\delta^2(\vec{q}_\perp +\vec{k}_{\perp} -\vec{k}^{\prime}_{\perp})\nonumber\\
 &\times \left[\left(c_1^q g_\perp^{\mu\nu}+i c_3^q \varepsilon_\perp^{\mu\nu}\right) f_1\mathcal{D}^h
+\left(c_3^q g_\perp^{\mu\nu}+i c_1^q \varepsilon_\perp^{\mu\nu}\right)f_1\mathcal{G}^h\right], \label{f:hadronich}
\end{align}
where
\begin{align}
  \mathcal{D}^h&=\mathcal{D}_1+\varepsilon_\perp^{h S} w_2 D^\perp_{1T} \nonumber\\
 &+h \cdot S_{LT}w_2 D^\perp_{1LT} +S^{hh}_{TT}w_{22}D^\perp_{1TT}, \label{f:Dh}\\
  \mathcal{G}^h&=\lambda_h G_{1L}+h \cdot S_{T}w_2 G^\perp_{1T}\nonumber\\
 &+\varepsilon_\perp^{h S_{LT}} w_2G^\perp_{1LT} +\tilde{S}^{hh}_{TT}w_{22}G^\perp_{1TT}. \label{f:Gh}
\end{align}
The $w$-factors in Eqs. (\ref{f:Dh}) and (\ref{f:Gh}) are defined as 
\begin{align}
  & w_2=\frac{\vec{h}\cdot \vec{k}^\prime_\perp}{M_h}, \\
  & w_{22}=\frac{2(\vec{h}\cdot \vec{k}^\prime_\perp)^2-(\vec{k}^{\prime }_\perp)^2}{M_h^2}.
\end{align}

\subsection{Cross section in the parton model}
The differential cross section is obtained by contracting the leptonic tensor and the hadronic tensor. First of all, we show the following contractions, 
\begin{align}
  & L_{\mu\nu}(c_1^qg_\perp^{\mu\nu}+ic_3^q\varepsilon_\perp^{\mu\nu})=-\frac{2Q^2}{y^2}T^q_0, \\
  & L_{\mu\nu}(c_3^qg_\perp^{\mu\nu}+ic_1^q\varepsilon_\perp^{\mu\nu})=-\frac{2Q^2}{y^2}T^q_1,
\end{align}
where
\begin{align}
  T^q_0 &=c_1^qA(y)-\lambda_\nu c_3^qC(y), \\
  T^q_1 &=c_3^qA(y)-\lambda_\nu c_1^qC(y).
\end{align}
The kinematic factors,  $A(y)=2-2y+y^2$, $C(y)=y(2-y)$.
Without showing the detailed complex calculation, we finally have
\begin{align}
  d\tilde{\sigma} = \frac{\alpha_{\rm em}^2U^{ij}}{2yz^2Q^2} A_W\mathcal{C} &\bigg\{T^q_0 f_1\Big(\mathcal{D}_1 -|S_T|\sin(\phi_h-\phi_S)w_2D^\perp_{1T} \nonumber \\
  &-|S_{LT}|\cos(\phi_h-\phi_{LT})w_2D^\perp_{1LT}\nonumber \\
  &+|S_{TT}|\cos(2\phi_h-2\phi_{TT})w_{22}D^\perp_{1TT}\Big)\nonumber\\
  +& T^q_1 f_1\Big(\lambda_h G_{1L}-|S_T|\cos(\phi_h-\phi_S)w_2 G^\perp_{1T}\nonumber \\
  & -|S_{LT}|\sin(\phi_h-\phi_{LT})w_2 G^\perp_{1LT}\nonumber\\
  &-|S_{TT}|\sin(2\phi_h-2\phi_{TT})w_{22}G^\perp_{1TT}\Big)\bigg\}, \label{f:crossfinal}
\end{align}
where $d\tilde{\sigma}= d\sigma/dxdyd\psi dzd^2\vec{p}_{h\perp}$ is used for convenience. $U^{ij}$ is the CKM matrix elements, $i$ denotes the flavor of the quark associated with the distribution function and  $j$ denotes the flavor of the quark associated with the fragmentation function.  The convolution is 
\begin{align}
  \mathcal{C}[w_i f_1D_1]=& \int d^2 \vec k_\perp d^2 \vec k_\perp^{\prime}\delta^2(\vec{k}_\perp^\prime - \vec{k}_\perp-\vec{q}_\perp)\nonumber\\
  & \times w_i f_1(x, k_\perp)D_1(z, k_\perp').
\end{align}

 We note that the flavor summation is understood for the total cross section. In the neutrino nucleus scattering, distribution functions are related to $d, s, \bar{u}, \cdots$ quarks while in the anti-neutrino nucleus scattering, distribution functions are related to $u, \bar{d}, \bar{s}, \cdots$ quarks. We only consider light flavors in this paper. 

\section{Measurable quantities}\label{sec:yielda}

\subsection{Azimuthal asymmetry}

According to the Trento conventions \cite{Bacchetta:2004jz}, the single spin asymmetry for the transverse polarization is written as
\begin{align}
  A_T=\frac{d\tilde{\sigma}(\phi_h,\phi_i)-d\tilde{\sigma}(\phi_h,\phi_i+\pi)}{d\tilde{\sigma}(\phi_h,\phi_i)+d\tilde{\sigma}(\phi_h,\phi_i+\pi)}.
\end{align}
The index $i$ can be $S, LT, TT$. The associated azimuthal asymmetry is therefore defined as, 
\begin{align}
  &\langle\sin(\phi_h-\phi_S)\rangle \nonumber\\
  &=\frac{\int \left[d\tilde{\sigma}(\phi_h,\phi_S)-d\tilde{\sigma}(\phi_h,\phi_S+\pi)\right]\sin(\phi_h-\phi_S)d\phi_h d\phi_S}{\int \left[d\tilde{\sigma}(\phi_h,\phi_S)+d\tilde{\sigma}(\phi_h,\phi_S+\pi)\right]d\phi_h d\phi_S}.
\end{align}
Other kinds of azimuthal asymmetries can be defined in a similar way. We do not show them here for simplicity. We list six kinds of asymmetries obtained from Eq. (\ref{f:crossfinal}) here,
\begin{align}
  & \langle\sin(\phi_h-\phi_S)\rangle =-\frac{1}{2}\frac{T_0^q\mathcal{C}\left[w_2 f_1 D^\perp_{1T} \right]}{T_0^q\mathcal{C}\left[ f_{1}D_1\right]}, \\
  & \langle\cos(\phi_h-\phi_S)\rangle =-\frac{1}{2}\frac{T_1^q\mathcal{C}\left[w_2 f_1 G^\perp_{1T} \right]}{T_0^q\mathcal{C}\left[ f_{1}D_1\right]}, \\
  & \langle\sin(\phi_h-\phi_{LT})\rangle =-\frac{1}{2}\frac{T_1^q\mathcal{C}\left[ w_2 f_1 G^\perp_{1LT} \right]}{T_0^q\mathcal{C}\left[ f_{1}D_1\right]}, \\
  & \langle\cos(\phi_h-\phi_{LT})\rangle =-\frac{1}{2}\frac{T_0^q \mathcal{C}\left[w_2 f_1 D^\perp_{1LT} \right]}{T_0^q\mathcal{C}\left[ f_{1}D_1\right]}, \\
  & \langle\sin(2\phi_h-2\phi_{TT})\rangle =-\frac{1}{2}\frac{T_1^q\mathcal{C}\left[w_{22} f_1 G^\perp_{1TT} \right]}{T_0^q\mathcal{C}\left[ f_{1}D_1\right]}, \\
  & \langle\cos(2\phi_h-2\phi_{TT})\rangle=\frac{1}{2}\frac{T_0^q\mathcal{C}\left[w_{22} f_1 D^\perp_{1TT} \right]}{T_0^q\mathcal{C}\left[ f_{1}D_1\right]}.
\end{align}
These asymmetries can be measured to extract corresponding nuclear PDF and FFs. Among these FFs,  $D^\perp_{1T} $ is important and known as Sivers type FF. It describes a unpolarized quark fragmenting into a transversely polarized hadron.

\subsection{Yield asymmetry of charged pion}

Equation (\ref{f:crossfinal}) is the complete differential cross section of the neutrino and/or anti-neutrino nucleus scattering for the production of the vector meson. It can be reduced to the differential cross section for the production of spinor particles by setting $S_{LL}=|S_{LT}|=|S_{TT}|=0$, for the production of pseudo scalar particles by further setting $\lambda_h=|S_T|=0$.
In this part, we consider the production of charged pion because the relevant FFs are available. 
To be explicit, we write down differential cross sections for the positively charged pion produced in the  neutrino nucleus scattering and for the negatively charged pion produced in the anti-neutrino nucleus scattering,
\begin{align}
  d\tilde{\sigma}_{\nu A}(\pi^+)=&\frac{\alpha^2_{\rm em}}{yz^2Q^2}A_WT_0\mathcal{C}\bigg\{f_1^d D_1^u(\pi^+) U^{ud}+f_1^{\bar{u}} D_1^{\bar{d}}(\pi^+)U^{ud}  \nonumber \\
  &+f_1^{\bar{u}} D_1^{\bar{s}}(\pi^+)U^{us}+f_1^s D_1^u(\pi^+)U^{us}\bigg\}, \label{f:crosszh} \\
  d\tilde{\sigma}_{\bar\nu A}(\pi^-)=&\frac{\alpha^2_{\rm em}}{yz^2Q^2}A_Wt_0\mathcal{C}\bigg\{f_1^u  D_1^d(\pi^-) U^{ud}+f_1^u  D_1^s(\pi^-) U^{us} \nonumber \\
  &+f_1^{\bar{d}} D_1^{\bar{u}}(\pi^-)U^{ud}+f_1^{\bar{s}} D_1^{\bar{u}}(\pi^-) U^{us}\bigg\}. \label{f:crossf}
\end{align}
Superscripts denote quark flavors, $U^{ij}$ are CKM matrix elements. In this paper, we use $U^{ud}=0.97367$ and $U^{us}=0.22431$ for numerical estimates \cite{ParticleDataGroup:2024cfk}.
Two $T$ functions are defined as
\begin{align}
  T_0&=A(y)+C(y), \\
  t_0&=A(y)-C(y).
\end{align}

We introduce the yield asymmetry between the productions of the positively charged pion  and the negatively charged pion, 
\begin{align}
  A^{\pi} =\frac{d\tilde{\sigma}_{\nu N}(\pi^+)-d\tilde{\sigma}_{\bar\nu N}(\pi^-)}{d\tilde{\sigma}_{\nu N}(\pi^+)+d\tilde{\sigma}_{\bar\nu N}(\pi^-)}, \label{f:asymmetry}
\end{align}
and obtain
\begin{align}
  A^{\pi}=\frac{A(y)F^\pi_M+C(y)F^\pi_P}{A(y)F^\pi_P+C(y)F^\pi_M}\label{f:difference}
\end{align}
by using Eqs. (\ref{f:crosszh}) and (\ref{f:crossf}). $F^\pi_M$ and $F^\pi_P$ are defined as
\begin{align}
  F_M^\pi  =&\mathcal{C}\Big[f_1^d D_1^u(\pi^+)-f_1^u D_1^d(\pi^-)\nonumber \\
  &+f_1^{\bar{u}} D_1^{\bar{d}}(\pi^+)-f_1^{\bar{d}}  D_1^{\bar{u}}(\pi^-)\Big] U^{ud}  \nonumber \\
   +&\mathcal{C}\Big[f_1^s D_1^u(\pi^+)-f_1^{\bar{s}} D_1^{\bar{u}}(\pi^-)\nonumber \\
  &+f_1^{\bar{u}} D_1^{\bar{s}}(\pi^+)- f_1^u  D_1^s(\pi^-)\Big]U^{us}, \label{f:fm}\\
  F_P^\pi  =&\mathcal{C}\Big[f_1^d D_1^u(\pi^+)+f_1^u D_1^d(\pi^-)\nonumber \\
  &+f_1^{\bar{u}} D_1^{\bar{d}}(\pi^+)+f_1^{\bar{d}}  D_1^{\bar{u}}(\pi^-)\Big] U^{ud}  \nonumber \\
   +&\mathcal{C}\Big[f_1^s D_1^u(\pi^+)+f_1^{\bar{s}} D_1^{\bar{u}}(\pi^-)\nonumber \\
  &+f_1^{\bar{u}} D_1^{\bar{s}}(\pi^+)+ f_1^u  D_1^s(\pi^-)\Big]U^{us}. \label{f:fp}
\end{align}

 Equations (\ref{f:fm}) and (\ref{f:fp}) can be simplified by using isospin symmetry. For FFs of the positively charged pion  and the negatively charged pion, we have
\begin{align}
  & D_1^u(\pi^+)=D_1^d(\pi^-), \\
  & D_1^{\bar{d}}(\pi^+)=D_1^{\bar{u}}(\pi^-).
\end{align}
The distribution functions of the up quark and that of the down quark also satisfy, 
\begin{align}
  & f_1^u=f_1^d, \\
  & f_1^{\bar{u}}=f_1^{\bar{d}},
\end{align}
since we are considering the SIDIS for isoscalar nuclei. $F^\pi_M$ and $F^\pi_P$ can therefore be rewritten as
\begin{align}
  F_{M,iso}^\pi  =&\mathcal{C}\Big[f_1^s D_1^u(\pi^+)-f_1^{\bar{s}} D_1^{\bar{u}}(\pi^-)\nonumber \\
  &+f_1^{\bar{u}} D_1^{\bar{s}}(\pi^+)- f_1^u  D_1^s(\pi^-)\Big]U^{us}, \label{f:fmiso}\\
  F_{P,iso}^\pi  =&2\mathcal{C}\Big[f_1^d D_1^u(\pi^+)+f_1^{\bar{u}} D_1^{\bar{d}}(\pi^+)\Big] U^{ud}  \nonumber \\
   +&\mathcal{C}\Big[f_1^s D_1^u(\pi^+)+f_1^{\bar{s}} D_1^{\bar{u}}(\pi^-)\nonumber \\
  &+f_1^{\bar{u}} D_1^{\bar{s}}(\pi^+)+ f_1^u  D_1^s(\pi^-)\Big]U^{us}. \label{f:fpiso}
\end{align}

We can further neglect contributions from sea quark distribution functions to obtain 
\begin{align}
  F^\pi_{M,V}=&-\mathcal{C}\left[f_1^u D_1^s(\pi^-)\right]U^{us},  \label{f:fmsea}\\
  F^\pi_{P,V}=&2\mathcal{C}\left[f_1^d D_1^u(\pi^+)\right]U^{ud}+ \mathcal{C}\left[f_1^u D_1^s(\pi^-)\right]U^{us}.  \label{f:fpsea}
\end{align}
and neglect contributions from disfavored FFs to obtain
\begin{align}
  F_{M,F}^\pi  =&\mathcal{C}\Big[f_1^s D_1^u(\pi^+)-f_1^{\bar{s}} D_1^{\bar{u}}(\pi^-)\Big]U^{us}, \label{f:fmisoR}\\
  F_{P,F}^\pi  =&2\mathcal{C}\Big[f_1^d D_1^u(\pi^+)+f_1^{\bar{u}} D_1^{\bar{d}}(\pi^+)\Big] U^{ud}  \nonumber \\
   &+\mathcal{C}\Big[f_1^s D_1^u(\pi^+)+f_1^{\bar{s}} D_1^{\bar{u}}(\pi^-)\Big]U^{us}. \label{f:fpisoR}
\end{align}
Note, Eqs. (\ref{f:fmsea})-(\ref{f:fpisoR}) are obtained from Eqs. (\ref{f:fmiso}) and (\ref{f:fpiso}).

\begin{figure}
  \centering
  \includegraphics[width=0.65\linewidth]{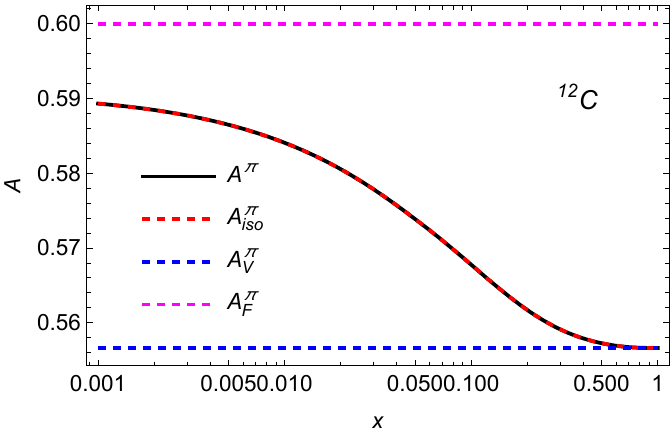}
  \caption{Numerical estimates of the asymmetry $A^\pi$ with respect to fraction $x$. The kinematic factor y is taken as $y=0.5$ and the fraction $z$ is taken as $z=0.2$. The solid black lines show estimates of $A^\pi$ without any approximation.}\label{fig:Apionx}
\end{figure}
\begin{figure}
  \centering
  \includegraphics[width=0.65\linewidth]{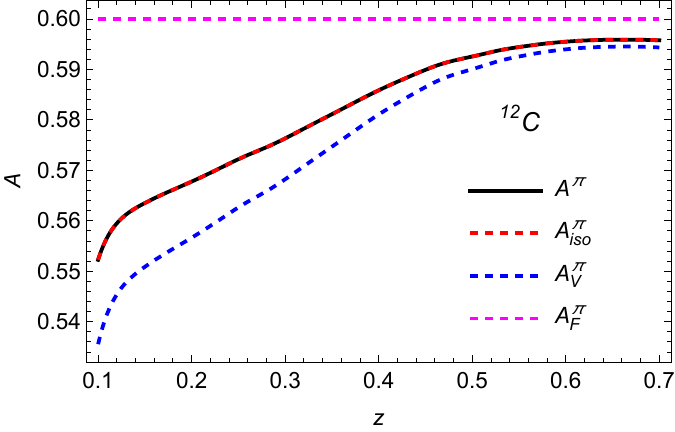}
  \caption{Numerical estimates of the asymmetry $A^\pi$ with respect to fraction $z$. The kinematic factor y is taken as $y=0.5$ and the fraction $x$ is taken as $x=0.1$. The solid black lines show estimates of $A^\pi$ without any approximation.}\label{fig:Apionz}
\end{figure}

Similar to Eq. (\ref{f:difference}), we can define $A_{iso}^\pi$, $A^\pi_V$ and $A^\pi_F$ by using $(F_{M,iso}^\pi, F_{P,iso}^\pi)$, $(F_{M,V}^\pi, F_{P,V}^\pi)$ and $(F_{M,F}^\pi, F_{P,F}^\pi)$. 
\begin{align}
  & A_{iso}^{\pi}=\frac{A(y)F^\pi_{M,iso}+C(y)F^\pi_{P,iso}}{A(y)F^\pi_{P,iso}+C(y)F^\pi_{M,iso}}, \label{f:differenceiso} \\
  & A_{V}^{\pi}=\frac{A(y)F^\pi_{M,V}+C(y)F^\pi_{P,V}}{A(y)F^\pi_{P,V}+C(y)F^\pi_{M,V}}, \label{f:differenceV} \\
  & A_{F}^{\pi}=\frac{A(y)F^\pi_{M,F}+C(y)F^\pi_{P,F}}{A(y)F^\pi_{P,F}+C(y)F^\pi_{M,F}}. \label{f:differenceF} 
\end{align}
In order to see the difference between $A^\pi, A^\pi_{iso}, A^\pi_V$ and $A^\pi_F$, we present numerical estimates in Fig. \ref{fig:Apionx} and Fig. \ref{fig:Apionz}. We take the Gaussian ansatz for the parametrization of the nuclear PDF $f_1$ and the FF $D_1$, 
\begin{align}
  & f_1^q(x, k_\perp) =f_1^q(x) \frac{1}{\pi \Delta^2}e^{-\vec{k}_\perp^2/\Delta^2}, \label{f:f1xperp}\\
  & D_1^q(z, k'_\perp) =D_1^q(z) \frac{1}{\pi \Delta_h^2}e^{-\vec{k}_\perp^{\prime 2}/\Delta_h^2}. \label{f:D1xperp}
\end{align}
By considering isospin symmetry, we here use $\Delta_u^2= \Delta_d^2=0.34$ GeV$^2$, $\Delta_{\bar{u}}^2= \Delta_{\bar{d}}^2=0.63$ GeV$^2$ and $\Delta_s^2=\Delta_{\bar{s}}^2=0.22$ GeV$^2$ \cite{Anselmino:2005nn,Signori:2013mda,Anselmino:2013lza,Cammarota:2020qcw,Bacchetta:2022awv,Bacchetta:2024qre} for light quarks, and $\Delta_h^2=0.17$ GeV$^2$ \cite{Bacchetta:2024qre,Callos:2020qtu} for light hadrons. 
The nuclear PDFs $f_1(x)$ are taken from \cite{Eskola:2021nhw,Hou:2019efy} while FFs are taken from \cite{Bertone:2017tyb}. The target nucleus considered here is carbon-12 ($^{12}C$), which has the same number of neutrons and protons. Numerical estimates are shown at $Q=5$ GeV and the kinematic factor $y$ is taken as $y=0.5$. The solid black lines show estimates of $A^\pi$ without any approximation while dashed colored lines show estimates of $A^\pi$ for different approximations.  
In Fig. \ref{fig:Apionx}, asymmetries are shown when $z=0.2$ while in Fig. \ref{fig:Apionz},  asymmetries are show when $x=0.1$. Because there are large uncertainties of FFs in the region of large or small $z$, we only consider the region of $(0.1, 0.7)$. 


From Fig. \ref{fig:Apionx} and Fig. \ref{fig:Apionz} we also notice that sea quark distribution functions and disfavored FFs have significant influence on measurable quantities. They should not be neglected in considering fragmentation processes. It is interesting to further consider $A^\pi_V$ and $A^\pi_F$ which are respectively defined by  $(F_{M,V}^\pi, F_{P,V}^\pi)$ and $(F_{M,F}^\pi, F_{P,F}^\pi)$. They exhibit very specific behaviors. According to our assumption, $A^\pi_V$ can be written as 
\begin{align}
  A^\pi_V = \frac{2C(y) D_1^{u,\pi^+}(z) +[C(y)-A(y)] D_1^{s,\pi^-}(z)}{2A(y)D_1^{u,\pi^+}(z)+[A(y)-C(y)]D_1^{s,\pi^-}(z)}, \label{f:ApiV}
\end{align}
due to $ f_1^u(x)=f_1^d(x)$ in the $N=Z$ nuclei.  $A^\pi_V$ is therefore only determined by FFs and is a constant when $y=0.5$ and  $z=0.2$ in Fig. \ref{fig:Apionx}. 
Similarly, $A^\pi_F$ can be written as 
\begin{align}
  A^\pi_F = \frac{C(y)}{A(y)} \label{f:ApiF}
\end{align}
as long as $f_1^s(x)=f_1^{\bar{s}}(x)$ is satisfied, and it therefore does not depend on fractions $x$ and $z$. This is the reason why we see $A^\pi_F(y=0.5) =0.6$ in these two figures and it always stays the same.  However, the violation of Eq. (\ref{f:ApiF}) would indicate the asymmetry between strange quark and anti-strange quark in the $N=Z$ nuclei. 

\begin{figure}
  \centering
  \includegraphics[width=0.65\linewidth]{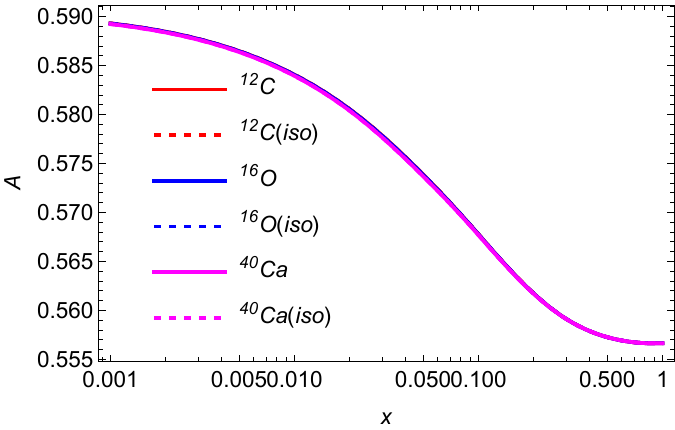}
  \caption{Numerical estimates of the asymmetry $A^\pi$ in SIDIS of $^{12}C$, $^{16}O$ and $^{40}Ca$. The kinematic factor y is taken as $y=0.5$ and the fraction $z$ is taken as $z=0.2$.}\label{fig:Acocax}
\end{figure}
\begin{figure}
  \centering
  \includegraphics[width=0.65\linewidth]{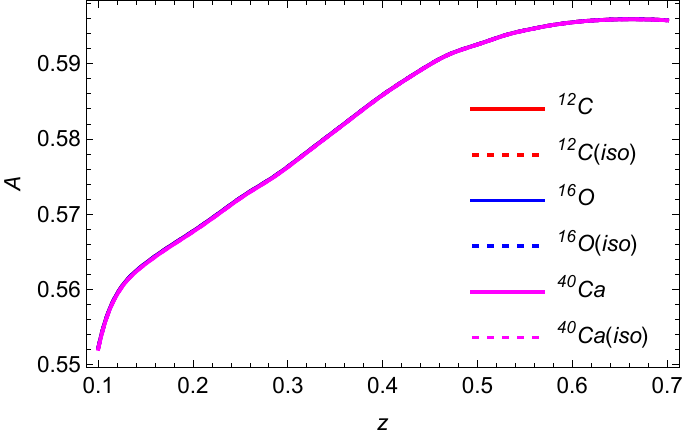}
  \caption{Numerical estimates of the asymmetry $A^\pi$ in SIDIS of  $^{12}C$, $^{16}O$ and $^{40}Ca$. The kinematic factor y is taken as $y=0.5$ and the fraction $x$ is taken as $x=0.1$.}\label{fig:Acocaz}
\end{figure}

Furthermore, we show the numerical results of $A^\pi$ and $A^\pi_{iso}$ in the SIDIS of  $^{12}C$, $^{16}O$ and $^{40}Ca$ in Fig. \ref{fig:Acocax} and Fig. \ref{fig:Acocaz}.  We notice that the yield asymmetry appears to be independent of the type of target nucleus as long as $N=Z$ is satisfied. To test this conclusion, we introduce the asymmetry ratio $R_A(C)$,  $R_A(O)$and  $R_A(Ca)$, see Appendix \ref{app:yield}.  Numerical estimates show that they vary by less than $0.0005$. At this level of precision, we can conclude that yield asymmetry is universal in charged current neutrino and anti-neutrino SIDIS of isoscalar nuclei. However, the universality of the yield asymmetry is broken in the scattering of other nuclei, see Appendix \ref{app:al}. 


\section{Summary}\label{sec:summary}

Parton distribution functions and FFs are important quantities in describing high energy reactions. To meet the needs of future high-energy neutrino experiments, we calculate the charged current semi-inclusive deeply inelastic neutrino and anti-neutrino nucleus scattering of $N=Z$ nuclei in this paper to study nuclear distribution functions. 
In order to obtain the explicit expression of the differential cross section, we calculate the neutrino and anti-neutrino nucleus scattering with the same formalism used in the SIDIS of nucleon. In this case, the differential cross section would be given in terms of convolutions of nuclear PDFs and FFs. According to calculations, we find that only chiral-even terms survive and chiral-odd terms vanish. In addition to presenting a set of azimuthal asymmetries, we further introduce a measurable quantity, the yield asymmetry of the produced hadrons, to study nuclear PDFs. In the Gaussian ansatz framework, numerical estimates show that the yield asymmetry appears to be independent of the type of target nucleus if is has the same number of neutrons and protons. This indicates that we can study the phenomenology by measuring the isospin symmetric particles. Numerical estimates also show that sea quark distribution functions and disfavored fragmentation functions have significant influence on measurable quantities.

Two issues that have been ignored in this paper need to be reconsidered. The first one is hidden in Eq. (\ref{f:pplus}) where the nucleus mass has been neglected. Nucleus usually has a large mass and it will break definitions in Eq. (\ref{f:sidisvar}). In this case, new definitions of these standard variables and new parameterizations of momenta should be considered further.  Reference \cite{Ruiz:2023ozv} argued that $Q^2/v$ is independent of the atomic number and perturbative techniques as well as the parton model still work. Standard variables can be rescaled. The second issue is the introduction of the distribution function ($f_1$) presented in Eq. (\ref{f:pdff1}). Systematic formalism of the inelastic scattering of nuclei at partonic level is still missing, therefore nuclear distribution functions obtained by decomposing the nulcear two-point correlation function without considering the degrees of freedom of nucleons need more research.

In this paper, we introduce the yield asymmetry to determine distribution functions. However, precise measurements require a high statistics, especially in neutrino reactions.  Recently, FASER (Forward Search Experiment) program at large hadron collider has reported the first  neutrino interaction candidates \cite{FASER:2021mtu}, the first measurement of the $\nu_e$ and $\nu_\mu$ interaction cross sections \cite{FASER:2024hoe} and the neutrino rate predictions \cite{FASER:2024ykc}. These recent discoveries have opened up new fields of neutrino research. We expect  it can potentially be used to measure these quantities.

\section*{ACKNOWLEDGEMENTS}

This work was supported by the National Natural Science Foundation of China (Grants No. 12405103, 12305106), Natural Science Foundation of Shandong Province (Grants No. ZR2021QA015, ZR2021QA040), and the Youth Innovation Technology Project of Higher School in Shandong Province (2023KJ146).

\begin{appendix}

\section{Yield asymmetry ratios}\label{app:yield}

\begin{figure}
  \centering
  \includegraphics[width=0.65\linewidth]{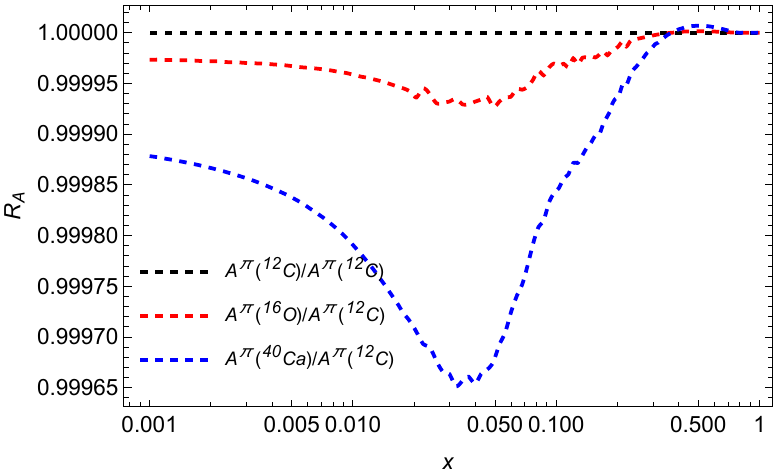}
  \caption{Numerical estimates of the asymmetry ratio of $R_A(C)$,  $R_A(O)$and  $R_A(Ca)$.}\label{fig:RA}
\end{figure}
\begin{figure}
  \centering
  \includegraphics[width=0.65\linewidth]{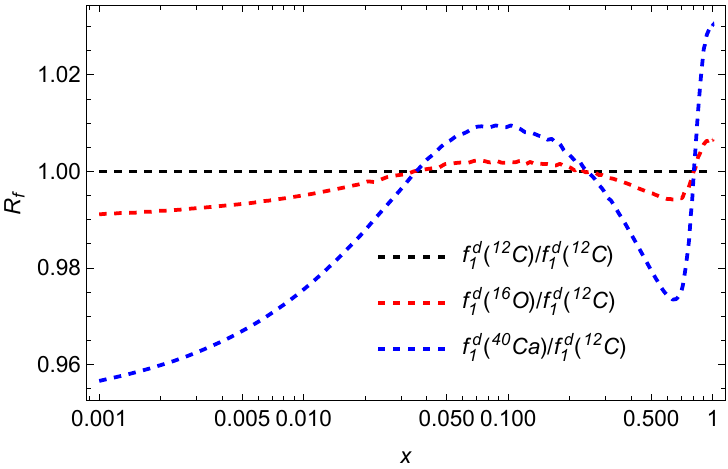}
  \caption{Numerical estimates of the ratio between nPDFs for $^{12}C$, $^{16}O$ and $^{40}Ca$.}\label{fig:Rf}
\end{figure}

 Yield asymmetry appears to be independent of the type of target nucleus provided it has the same number of neutrons and protons. In order to see the universality of the yield asymmetry, we introduce the asymmetry ratio defined by the following form, 
\begin{align}
  R_A(A_2)= \frac{A^\pi(A_2)}{A^\pi(C)}, 
\end{align}
where $A_2=C, O, Ca$.  We present numerical estimates in Fig. \ref{fig:RA}, which shows that ratios are approximately equal to $1$. 

In order to see the effect of nPDFs on the ratio, we also introduce the ratio between distribution functions, 
\begin{align}
  R_f (A_2)=\frac{f_1(A_2)}{f_1(C)}. 
\end{align}
It should be noted that nPDF for other $N=Z>20$ nuclei are not available. For these available ones, we present numerical estimates in Fig. \ref{fig:Rf}. We here only show estimates for the down quark. The results for the strange quark are similar. From Fig. \ref{fig:RA} and Fig. \ref{fig:Rf}, we notice that the differences between $R_f (A_2)$ are larger than  the differences between $R_A (A_2)$.

\section{Universality breaking}\label{app:al}

\begin{figure}[h]
  \centering
  \includegraphics[width=0.65\linewidth]{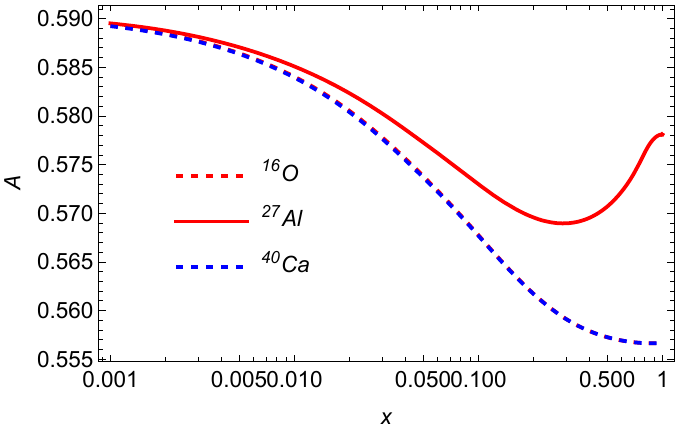}
  \caption{Numerical estimates of the asymmetry $A^\pi$ in SIDIS of $^{16}O$, $^{27}Al$ and $^{40}Ca$. The kinematic factor y is taken as $y=0.5$ and the fraction $z$ is taken as $z=0.2$.}\label{fig:Alx}
\end{figure}
\begin{figure}[h]
  \centering
  \includegraphics[width=0.65\linewidth]{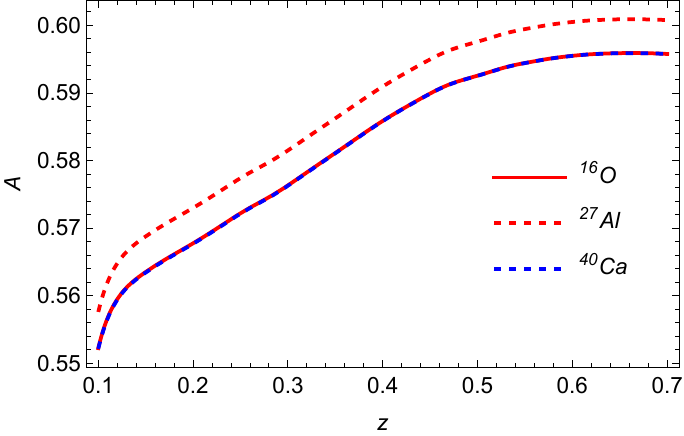}
  \caption{Numerical estimates of the asymmetry $A^\pi$ in SIDIS of  $^{16}O$, $^{27}Al$ and $^{40}Ca$. The kinematic factor y is taken as $y=0.5$ and the fraction $x$ is taken as $x=0.1$.}\label{fig:Alz}
\end{figure}

To verify that the universality of the yield asymmetry is broken in the scattering of $N\neq Z$ nuclei, we here show numerical estimates of $A^\pi$ in the SIDIS of $^{16}O$, $^{27}Al$ and $^{40}Ca$ in Fig. \ref{fig:Alx} and Fig. \ref{fig:Alz}.  For aluminum, $N=14, Z=13$, we can see that the universality of the yield asymmetry is broken.

\end{appendix}

\end{document}